\documentclass[conference]{IEEEtran}
\IEEEoverridecommandlockouts
\usepackage{cite}
\usepackage{amsmath,amssymb,amsfonts}
\usepackage{algorithmic}
\usepackage{graphicx}
\usepackage{textcomp}
\usepackage{xcolor}
\usepackage{subcaption}
\usepackage{adjustbox}
\def\BibTeX{{\rm B\kern-.05em{\sc i\kern-.025em b}\kern-.08em
    T\kern-.1667em\lower.7ex\hbox{E}\kern-.125emX}}
\begin{document}

\title{From Emulation to Mathematical – A More General Traffic Obfuscation Approach To Encounter Feature based Mobile App traffic Classification\\

}

\author{\IEEEauthorblockN{Lichun Gao}
\IEEEauthorblockA{\textit{Computer Science Department} \\
\textit{Worcester Polytechnic Institute}\\
Worcester, USA \\
lgao2@wpi.edu}
\and
\IEEEauthorblockN{Mingjie Zeng}
\IEEEauthorblockA{\textit{Computer Science Department} \\
\textit{Worcester Polytechnic Institute}\\
Worcester, USA \\
mzeng2@wpi.edu}
\and
\IEEEauthorblockN{Zhanhong Huang}
\IEEEauthorblockA{\textit{Computer Science Department} \\
\textit{Worcester Polytechnic Institute}\\
Worcester, USA \\
zhuang5@wpi.edu}
}

\maketitle



\textbf{\textit{Abstract}—The usage of the mobile app is unassailable in this digital era. While tons of data are generated daily, user privacy security concerns become an important issue. Nowadays, tons of techniques, such as machine learning and deep learning traffic classifiers, have been applied to analyze users' app traffic. These techniques allow the monitor to get the fingerprints of using apps while the user traffic is still encrypted, which raises a severe privacy issue. In order to fight against this type of data analysis, people have been researching obfuscation algorithms to confuse feature-based machine learning classifiers with data camouflage by modification on packet length distribution. The existing works achieve this goal by remapping traffic packet length distribution from the source app to the fake camouflage app. However, this solution suffers from its lack of scalability and flexibility in practical application since the method needs to pre-sample the target fake app's traffic before the use of traffic camouflage. In this paper, we proposed a practical solution by using a mathematical model to calculate the target distribution while maintaining at least 50\% accuracy drops on the performance of the AppScanner mobile traffic classifier and roughly 20\% overhead created during packet modification.
}   
\section{Introduction}
\subsection{Problem Description}
There has been tremendous growth in network applications in the last half decades. As one of the main players in this game, smartphone applications have become the main practice of digital interaction in people's life. According to TechCrunch\cite{b1}, in the second quarter of 2022, Android users in the US spend more than 4 hours daily on average on smartphone apps. While the average usage of mobile phones is increasing continuously, a huge amount of smart devices are getting connected through the Internet of Things (IoT) networks. As IoT devices reveal user behavior and transmit user personal information, applications used to manage these devices bridge our privacy to the Internet.
\par
With the explosive growth of network-based applications, a tremendous amount of data traffic is being created. According to App.ai\cite{b2}, 295 billion USD was spent on Mobile ads in 2021, with a 23\% growth compared to 2020. Data scientists are analyzing these traffic data to create huge benefits. Research has shown that traffic classification is a promising technology contributing to delineating security strategies, monitoring botnet propagation, and filtering traffic\cite{b3,b4,b5,b6}. Traffic classification can also support the control and management of resources in TCP/IP networks by designing QoS strategies \cite{b7}.
\par
The selection of applications is usually followed by users' habits, activities, health, and even more. By knowing the types of applications installed by a user, the data collector can infer much more information behind those data traffic. Different encryption techniques have been applied to protect user privacy to preserve traffic confidentiality.
\par
However, the increase in traffic classification implemented with machine learning methodology has diminished traditional techniques of encrypting data. Bu, Zhiyong, et al.\cite{b8} proposes deep and parallel network-in-network models to improve accuracy in traffic characterization. Pham, Thai-Dien, et al.\cite{b9} implements deep graph Convolution Neural Networks (CNNs) to classify mobile applications specifically. Aceto, Giuseppe, et al. \cite{b10} introduces Markov Modeling to characterize Mobile-app traffic and predict data traffic. An attacker can build a classifier and sniff mobile-app protocols with these techniques easily, which infer that traffic analysis could result in a big issue for user privacy.
\par
In this research project, we aim to develop a mobile-app traffic mutation algorithm to protect users' privacy. Our goal is to provide a solution to fight against the intruder by monitoring app traffic and using statistical analysis to infer users' personal information. While maintaining the traffic data anonymity, we also need to ensure that the traffic overhead and keep computation cost remaining in a reasonable level. We also want to provide solid strategies for different types of heterogeneous apps to improve the algorithm's scalability and dynamics.

\subsection{Limitation of Current Approaches}
As the importance of mobile-app traffic security is being noticed, multiple studies on mobile-app traffic camouflage have been directed from academic communities; however, most of them still have limitations on some extent. Sengupta, Satadal, et al.\cite{b11} exploits current traffic classification methods and denotes the main types of features of traffic characteristics, which are Packet-sized based features, inter-packet arrival time (IAT, average (aka arithmetic mean) of the times between packets arriving at a host over a period. It is commonly referred to as delay) based features, and Bit-sequence based features. They also mentioned most of the classical traffic mutations focus on changes in packet-size-based features and IAT-based features. Salman, Ola, et al.\cite{b12} extends the traditional methods by using Generative Adversarial Network (GAN) to train an auto-encoder to edit the source traffic. However, either auto-encoded traffic or manually adjusted traffic would create considerable overheads. On the other hand, Chaddad, Louma, et al.\cite{b13} considers reducing overhead as their optimal goal while maintaining traffic security. They first record a packet size table for a target app traffic to which the source app traffic will be faking. Then the sources app packet-sized probability distribution will be adjusted to a similar distribution of target app traffic. During this process, they match the closest-size packets between two traffic packet size distributions to minimize the overheads (padding added). Similarly, they apply fragmentation if a longer packet needs to be shortened. But the downside of this method is also obvious. While an app's traffic is being mutated to another one's, it is also shown as a feature of the source app since the mutation logic wouldn't be changed in a short time. The intruder can soon realize that the traffic shaped like a target app is faked. Also, the application of their algorithm for different kinds of mobile apps will need to match the source and target one by one to achieve the best optimization, which is hard to implement in reality. Based on the quick review given for existing approaches for mobile-app traffic mutation, we derive that it is urgent to develop a general resource-saving method while de-featuring the traffic to make communications between smartphones and the Internet more secure. 
\subsection{Main Challenges}
Making traffic anonymous and encrypting while saving resources is an argument within the context of designing an encryption algorithm. Finding such a balance point to achieve the best performance is always hard, thus we need to treat this challenge carefully. while Salman, Ola, et al.\cite{b12} can implement GAN to get a reasonable optimal point, the actual sweet point would always depend on the user's application environment. For ease of experience, our algorithm may allow the developer to adjust the mutation level by tweaking parameters to achieve the algorithm dynamics and scalability.
\par
But as machine learning techniques have more actual practices in traffic classification, patterns of our algorithm will eventually be found. Our algorithm would be dismembered by stochastic analysis using deep learning. One way to tackle this problem is to allow the algorithm switches its mode or be re-coded after several times of operations, which is implemented by \cite{b12}. Another way is to add "noise" into mutated traffic to fight against stochastic data analysis. But this will go back to the first challenge we discuss, how much noise will be reasonable to make our algorithm survive longer? The answer is also similar to our first one: it depends.
\par
Furthermore, it is challenging to run the race with not just the elites in the machine learning industry but also our peers. While Chaddad, Louma, et al.\cite{b13} have already achieved a result by reducing classification accuracy from 91.1\% to 0.22\% with 11.86\% padding overhead and to 1.76\% with only 0.73\% overhead, there is not a lot of space that allows us to "push forward." Therefore, our main goal is to design the algorithm with more generosity, which is to lower the pre-processing cost of the algorithm application.
\par
Finally, building a lightweight framework running on mobile platforms while keeping the features above would also be a challenge. As cellular technologies grow exponentially in this half-decade, the amount of traffic that a smartphone generates at the same time will also increase explosively. Service Providers in the US are closing 3G cellular networks and putting the 4G, and 5G on the table \cite{b14}. And Intel and Broadcom just released their Wi-Fi 7 demo with transfer speeds of 5 Gbps, which is five times the Wi-Fi 6 \cite{b15}. On the contrary, mobile phone processors haven't grown much as the semiconductor industry's development no longer follow Moore's law. Snapdragon 8 Gen 1, the flagship processor on the Android Platform of Broadcom, only gets 8\% more on single core performance and 2\% more on the multi-cores, compared to the previous flagship Snapdragon 888 \cite{b16}. One of the solutions for this challenge is to include GPU calculation in the encoding process, which is much more efficient than the CPU process. However, our main goal of the project is not optimizing the smartphone platform, so lowering the computation cost would not be our priority before we have a solid achievement of our other targets.
\subsection{Summary of Evaluation \& Results}
To evaluate our obfuscation algorithm's performance, we first use packet-features traffic data from MIRAGE-2019 dataset \cite{b25}to train a Random Forest Classifier -- AppScanner\cite{b28}. With 75\% training and 25\% test ratio data separation, we validate our mutated packet effectiveness on the classifier. We shoot a 50\% classifier accuracy drop and 20\% overhead packet on average, which achieve a similar performance as peer's work but can still have improvement.
\subsection{Summary of Paper}
In this paper, we proposed a mathematical traffic obfuscation method to solve the feature-based mobile app traffic classification. In Section II, we explore the existing traffic classification techniques, methods, and models to find the solution for enhancing our algorithm. Section III describes the data set we used, how we implemented the model, and the details of our algorithm. In Section IV, we displayed our experiment results and expanded the observations. Then in Section V, we conclude our work and discuss the promising future work.   

\section{Literature Review}

\par
In this research project, we aim to develop a mobile-app traffic mutation algorithm to protect users’ privacy. Our goal is to provide a solution to fight against the intruder by monitoring app traffic and using statistical analysis to infer users’ personal information. While maintaining the anonymity of traffic data, we also need to ensure the traffic overhead and keep the computational cost at a reasonable level. We also want to provide solid strategies for different types of heterogeneous apps to improve the algorithm’s scalability and dynamics.

\subsection{Traffic security}
As the importance of mobile-app traffic security is being noticed, multiple studies on mobile-app traffic camouflage have been directed by academic communities. Moreover, traffic mutation techniques nowadays can be categorized into different groups. The common way is to mutate the traffic from the source applications to the target ones\cite{b17}. Chaddad et al.\cite{b17} proposed confusion models as the solution. Those confusion models can obfuscate packet length information leaked by mobile traffic and shape one traffic class to obscure its class features. However, it is hard to decide the cost of shaping the traffic. Further, another way of traffic mutation using the probabilistic distribution of packet sizes came up\cite{b18}. Now without another application’s model traffic, the success of traffic mutation could still be achieved by some features’ probabilistic distribution instead of resembling the traffic with another application. For example, in \cite{b19}, they choose the packet sizes as the traffic features and model the packet lengths probability distribution of the source and target applications. A security model will mutate the packet length of the source application to the target application, and the probability distribution of packet sizes is similar to bin probability. The involvement of the machine learning method is also essential in the implementation and improvement process. In \cite{b12}, they construct an unsupervised deep learning model to detect the mutated traffic and train an en-coder to modify the source traffic. But traditional techniques for traffic classifying do not work well for mobile apps due to the lack of unique signatures\cite{b20}. In \cite{b20}, they applied different features to the classification and experimented with obtaining the most distinctive features in the mobile apps’ traffic. In the following paragraphs, we will describe each proposed project separately.

\subsection{App Traffic Mutation}
Chaddad et al. present a methodology\cite{b17} for identifying mobile apps using traffic analysis and propose confusion models that obfuscate packet length information by shaping one class of app traffic to obscure its class features with minimum overhead. This method shapes an app's flows so they maximally look like flows generated by another app. They also assess the model's efficiency using different apps and against a recently published approach for mobile app classification. However, it is hard to tell the cost of shaping the traffic.

\subsection{Packet Camouflage in Traffic Analysis}
On the other hand, Chaddad, Louma, et al. \cite{b13} considers reducing overhead their optimal goal while maintaining traffic security. They first record a packet size table for a target app traffic to which the source app traffic will be faking. Then the sources app packet-sized probability distribution will be adjusted to a similar distribution of target app traffic. During this process, they match the closest-size packets between two traffic packet size distributions to minimize the overheads (padding added). Similarly, they apply fragmentation if a longer packet needs to be shortened. But the downside of this method is also obvious. While an app’s traffic is being mutated to another one’s, it is also shown as a feature of the source app since the mutation logic would not be changed in a short time. The intruder soon realizes that the traffic shaped like a target app is faked. Also, the application of their algorithm for different kinds of mobile apps will need to match the source and target one by one to achieve the best optimization, which is hard to implement in reality.
While Chaddad, Louma, et al. \cite{b13} have already achieved a result by reducing classification accuracy from 91.1\% to 0.22\% with 11.86\% padding overhead and to 1.76\% with only 0.73\% overhead, there is not much space that allows us to ”push forward.” Therefore, our main goal is to design the algorithm with more generosity, which is to lower the pre-processing cost of the algorithm application.
Finally, building a lightweight framework running on mobile platforms while keeping the above features would also be challenging. As cellular technologies grow exponentially in this half-decade, the amount of traffic that a smartphone generates at the same time will also increase explosively. However, our main goal of the project is not optimizing the smartphone platform, so lowering the computation cost would only be our priority after we achieve our other targets.

\subsection{Mobile Traffic Anonymization}
To achieve the security goal and protect the privacy information involved in the use of mobile app, recent research always solves this problem by mutating app traffic by resembling the traffic of another app. In \cite{b18}, Chaddad et al. develop a simpler and more scalable system to anonymize mobile app packet traffic without needing another app’s model traffic using the probabilistic distribution of packet sizes. They first propose a scheme that regenerates statistic modeling of app packet lengths and then use the regenerated packet lengths to mutate the incoming traffic.

\subsection{Traffic Classification Diversity}
Sengupta, Satadal, et al. \cite{b11} exploits current traffic classification methods and denote the main types of features of traffic characteristics, which are Packet-sized based features, IAT based features, and Bit-sequence based features. They also mentioned that most of the classical traffic mutations focus on changes in packet-size and IAT-based features.

\subsection{Autoencoder for Traffic Detection and Recovery}
Salman, Ola, et al. \cite{b12} extends the traditional methods by using a Generative Adversarial Network (GAN) to train an auto-encoder to edit the source traffic. However, either auto-encoded traffic or manually adjusted traffic would create considerable overheads.
Based on the quick review given for existing approaches for mobile-app traffic mutation, it is urgent to develop a general resource-saving method while de-featuring the traffic to make communications between smartphones and the Internet more secure.
Making traffic anonymous and encrypting while saving resources is an argument within the context of designing an encryption algorithm. Finding such a balance point to achieve the best performance takes time and effort.
While Salman, Ola, et al. \cite{b12} can implement GAN to get a reasonable optimal point, the actual sweet point would depend on the user’s application environment. For ease of experience, our algorithm may allow the developer to adjust the mutation level by tweaking parameters to achieve the algorithm dynamics and scalability.
However, as machine learning techniques have more actual practices in traffic classification, patterns of our algorithm will eventually be found. Our algorithm would be dismembered by stochastic analysis using deep learning. One way to tackle this problem is to allow the algorithm switches its mode or be re-coded after several times of operations, which is implemented by \cite{b12}.
Another way is to add ”noise” into mutated traffic to fight against stochastic data analysis. However, this will go back to the first challenge we discuss, how much noise will be reasonable to make our algorithm survive longer? The answer is also similar to our first one: it depends.

\subsection{Data Mining/Machine Learning in cyber security}
Cyber Security\cite{b4}, defined as the set of technologies and processes to protect computers, networks, programs, and unauthorized access, has been an important issue with the development of computer networks. The intrusion has external ones (attacks from outside the organization) and internal ones (attacks from within the organization). The system we build to help us detect the intrusion is called IDS (intrusion detection system). 
Multiple Machine learning and data mining methods have been raised, and some most often used methods can be concluded as follows. 

\subsubsection{Artificial Neural Network}
ANN helps us with the two aspects we mentioned earlier, misuse detection and anomaly-based detection. In Misuse Detection, Cannady\cite{b21} used ANNs as a multi-category classifier to detect misuse. In the stage of data processing, it will result in nine features, including protocol identifier, source port, destination port, source address, a destination address, ICMP type, ICMP code, raw data length, and raw data. In conclusion, an RMS of 0.070 can roughly be considered as 93\% accuracy for the testing phase; each packet will be recognized as either a normal or attack group. In Anomaly and Hybrid Detection, Lippmann and Cunningham\cite{b22} proposed a system based on keyword selection and ANN. The keyword has the input to a NN that provides the probability of attack; the second NN operates on the flagged instance as attacks. Both NN consisted of multi-layer without hidden units.

\subsubsection{Bayesian Network}
The Bayesian network is a probabilistic graphical model that resents relationships between the variables. In misuse detection, the system is proactive because the signatures will be taken out from the input and checked continuously against the various attack patterns. Livadas et al.\cite{b23} resolve the botnet traffic in Internet Relay Chat (IRC) traffic, using TCP-level data to generate the network streams or NetFlow data. In Anomaly and Hybrid Detection, when the platform receives TCP/IP packets, the network stack of the underlying operating system will process the packets. The network comes out with generating logs and system kernel calls.  

\subsection{Deep Learning/Machine Learning in Network Traffic}
Network Traffic classification\cite{b24} is widely used in various applications. In real life, most applications will choose to encrypt their network traffic and change their port numbers dynamically. Even in this case, Machine Learning(ML) and especially Deep Learning(DL)-based classifiers have demonstrated impressive performance in network traffic classification. 
Knowing that network traffic consists of bidirectional flows and packets,  Sadeghzadeh et al.\cite{b24} designed an ML-based classifier as a function that maps the input space to an output space. The input space can be considered into three categories: 1) packet classification, 2) flow content classification, and 3) flow time series classification. Each of them has different features to consider the packets and flows. Furthermore, the DL-based classifiers are mainly based on deep neural networks. In the network traffic study, Convolutional Neural Network (CNN), Recurrent Neural Network (RNN), and Stacked Denoising Autoencoders (SDAE) are the three main DNNS that we will focus on. 
From analyzing Sadeghzadeh et al.\cite{b24}'s study, we can tell that the robustness of DL-based network traffic classifiers is a crucial aspect. It is critical when we are using the Deep Learning method to solve network traffic issues.

\subsection{Deep Learning in Networking}
Deep Learning is essentially a sub-branch of ML\cite{b7}\cite{b3}, enabling the algorithm to make predictions, classifications, or decisions based on large-scale data. Compared to traditional machine learning, which relies heavily on features defined by human experts, the deep learning algorithm can extract knowledge from raw data, reducing the cost extensively. 
\subsubsection{Advantages of Deep Learning in Networking}
Machine learning in networking requires a high amount of domain experts’ knowledge to build the features; however, with the usage of deep learning, the learning process will not be required to build by human experts anymore. Deep learning is also capable of handling large amounts of data. Mobile networks have countless uses, generating high volumes of different types of data at a very fast pace. ML is not enough in this condition, and deep neural networks can benefit us by training with big data without model over-fitting. Moreover, deep learning in networking solves the problem of labeling, which is highly costly. Deep neural networks can learn compressive representations more efficiently and better deal with geometric data.  
\subsubsection{Disadvantages of Deep Learning in Networking}
In our study, we need to avoid the drawbacks that deep learning may bring. Firstly, it is harmful to those adversarial examples; the attacker-designed intrusion inputs may trick the model into generating errors. Hackers may exploit the weakness in Neural Network models and training processes to perform attacks that disrupt deep learning-based cyber-defense systems. Deep learning algorithms are large black boxes that we should be careful when using, and it relies heavily on data, so the input type of data will be an important factor in our project. 

\section{Methodology}
\subsection{Attack Model}
\par 
In our experiment model, we study the network traffic with two end hosts running the same application at the same time, one end is the mobile phone, and the other one is the data server. The communication between two hosts implements encryption within a specific channel. Even with the encryption data, the attackers do not have knowledge about the raw data, but the features of packet flow(i.e., source/destination IP, source/destination port, packet size) could be used for classification. While monitoring is operating, traffic analysis attacks can be implemented as a classifier that tries to recognize the running application within the encrypted communication.
\par
Figure 1 demonstrates our attack model, in which an adversary aims to determine a victim's online activity. In this case, we assume that all the activities of program-based and data-based user applications appear on the same network, excluding the situation where users use WiFi and cellular networks for communication simultaneously. The direction of data forwarding contains both uploading and downloading. And the types of applications also could be many, including chatting, online gaming, video streaming, web page browsing, voice-over-IP, etc. In this experiment, we assume users perform the activities mentioned above by running only one mobile application at a time. For the attacker side, we assume network traffic fingerprinting is employed to sniff the encrypted data between the users and the server. A sniffer software adopted by the adversary side (e.g., Wireshark) to collect exchanged data but does not have any knowledge about the software identity and the encrypted logistics (e.g., key, MD5, scheme, etc.) We also assume that the sniffer can have full access and traces of the communication tunnel, such as physical access to the servers, in the case that the sniffer might be the maintainer of the server. But since we assume the adversary doesn't have any knowledge about the encrypted schemes, the sniffer can not decrypt the collected packets and has to implement machine learning to map down the traffic fingerprints with the flow features in order to match the application user using, even in the circumstance that the payloads are not accessible.
\par
Since the attacker is not able to read the packet directly, he might inspect the side channel information (IP packet headers) that is the companion to the encrypted data traffic. The attacker will use a classifier that has been pre-trained for the number of applications and then try to match the captured traffic from users to infer the app being used. We assume that the data capture is passive and not detectable by users so that the attack can continue as long as the attacker wants. Our proposed method intends to prevent the attacker from successfully distinguishing the app's detail during the process mentioned above. In short, we aim to prevent malicious identification of encrypted data through a data capture analysis.
\par
Figure 2 shows our demonstration of the traffic classification attack. We use this attack model to test our obfuscation model in the Empirical Result Section. We consider users to use a mobile phone while using the app connecting to the Internet. The data stream created by the end is end-to-end encrypted. The adversary captures the data and traces side channel information and tries to classify the app being used by the users.
\par
After the capture and analysis, the attacker would deploy supervised machine learning algorithms with a training and validation phase. While training, the supervised learning model was given $(X_i,y_i)$ where each $X_i$ is a vector of features, and $y_i$ is a ground truth label. And then, for each group of data, the classifier is given a vector Z, which will return an estimation label to indicate the precise mobile app that creates the data flow. In our experiments, $X_i$ contains information on the packet lengths, IAT, and the direction of packet flow(upload or download) of the captured encrypted data stream.
\par
In summary, the current existing implementations of network-based mobile apps have given limited security assurances against analysis or the need to choose a specific target app for camouflage, which created a lack of generosity in implementations. Our proposed solution needs to be efficient compared to the traditional obfuscation method, costing low computational and storage overhead. Besides, the algorithm also needs to be able to plug and play and doesn't need to set up a preset before the implementation.

\begin{figure}
    \centering
    \begin{subfigure}[b]{0.4\textwidth}
        \centering
        \includegraphics[scale = 0.25]{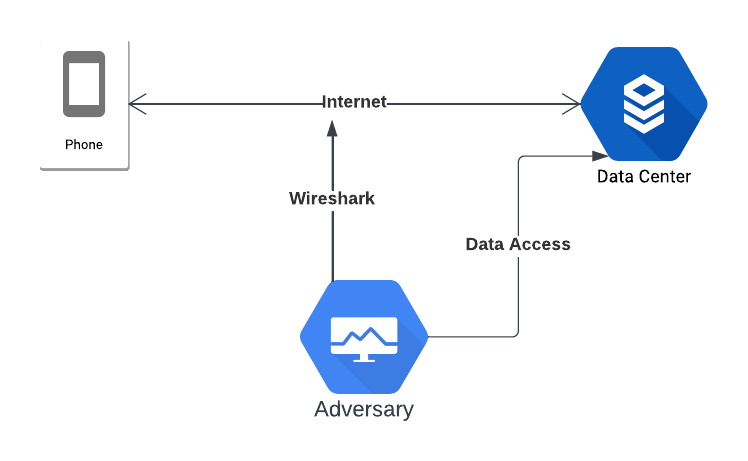}
        \caption{Threat Model}
        \label{fig:threat model}
    \end{subfigure}
    \hfill
    \begin{subfigure}[b]{0.4\textwidth}
        \centering
        \includegraphics[scale = 0.2]{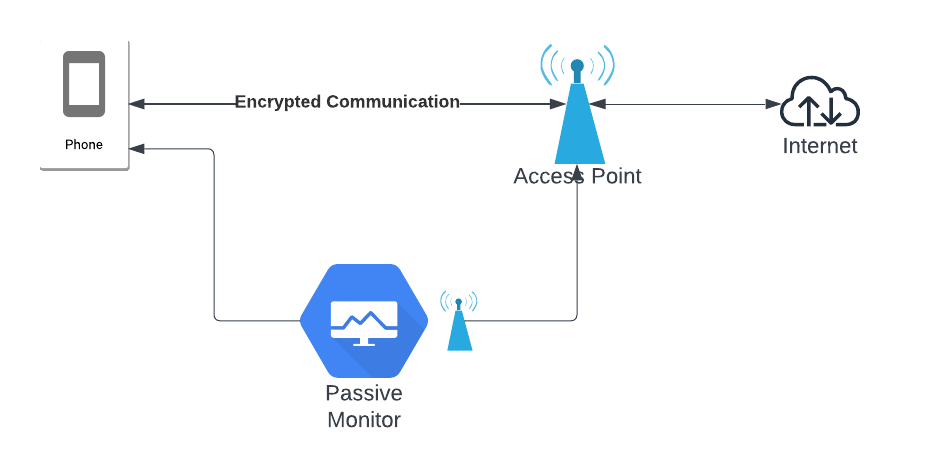}
        \caption{Mobile Traffic Classification Attack}
        \label{fig: Traffic Attack Model}
    \end{subfigure}
\end{figure}

\subsection{Proposed Frameworks}\label{framworks}
Our algorithm is aimed to distract the network fingerprint attacks based on statistical traffic analysis by confusing attackers with the camouflage app's traffic. Just like the literature mentioned, the main characteristics that currently traffic analysis traces for encrypted data streams are the packet lengths and the IAT of continuous packets. Our mutating target will focus on adjusting the packet size since modifying IAT will cause a delay issue. While it could take shorter the IAT to confuse the classifier, the space for modification would be varied and not very practical.
\par
In the following section, we will discuss our algorithm, which mutates the intakes packets. On the implementation scope, our algorithm acts like a proxy between applications and the network layer. During the proxy-like process, we will set the flag for each packet to show if they are padded or fragmented. If a packet is cut and loses the endpoint, a notation padding would also be attached for recovering the original packets.

\subsubsection{Define probabilistic distribution matching model}
In our algorithm, we adjust the length of each packet in the flows, which is created by the source app to be defended, so that the packet size probability distribution becomes different from the original packet sizes of the source app. The first step is to choose a preferred packet size distribution. We define our packet length probabilistic model as a skew-normal distribution based on the nature of the packet length distribution in a bitflow, which also shows a skewed normal distribution. In our probabilistic matching model, where $x$ represents the endpoint of the position calculated by the integral of desired function equal to the summation of the packet size in a single flow. In the pre-skewed normal distribution (PDF, eq.(\ref{eqn1})\cite{b29}) $\phi(x)$, $\sigma$ represents the standard deviation of the distribution, and $\mu$ represents the mean of the distribution. Then cumulative distribution function (CDF, eq.(\ref{eqn2})\cite{b29}) $\Phi(x)$ is calculated by integral the $\phi(x)$. Finally, we calculated the skew distribution (PDF, eq.(\ref{eqn3})\cite{b29}) by implementing skewness parameter $\alpha$, where the distribution is skewed left when $\alpha$ is positive and skewed right when $\alpha$ is negative. The parameters shown above could be predefined before the obfuscation process starts or could be set to bounded variable changed based on timestamp $t$.
\begin{equation}
\label{eqn1}
    \phi(x) = {1\over{\sigma\sqrt{2\pi}}}e^{{-1\over2}{{x-\mu}\over \sigma}^2}
\end{equation}
\begin{equation}
\label{eqn2}
    \Phi(x) = \int_{0}^{x} \phi(x) \,dx \
\end{equation}
\begin{equation}
\label{eqn3}
    f_a(x)={{\phi(x)\Phi(\alpha x)}\over{\Phi(0)}}
\end{equation}

\subsubsection{Adjust the source packet length distribution}
\par
Once the target packet size probability distribution is decided, we need to modify the src packet size distribution while minimizing the modification. For each incoming packet $p$ of source app of size $L_p$ having a probability $P_p$, we need to minimize the $P_p$ between two distributions for each $L_P$, which is shown in eq.(\ref{eqn4}).
\begin{equation}
\label{eqn4}
   \textbf{Min}|{P_p - {P_p}^`}| \textbf{ for each } L_p  
\end{equation}

One intuitive way to achieve this modification is by adding padding along each packet so that the distribution can be matched. However, quite an amount of overhead data would be created by modification due to the padding if two distributions have great differences in shape. The optimization problem could be formulated as eq.(\ref{eqn5}), where ${L_p}^`$ is the packet size for ${P_p}^`$.

\begin{equation}
\label{eqn5}
    \textbf{Min}\sum_{x=0}^{\infty} \sqrt{({P_p - {P_p}^`})^2\times ({L_p - {L_p}^`})^2 }
\end{equation}

\par To minimize the overhead, we introduce two techniques while modifying the packet size: fragmenting and stacking. Fragmenting is used for cutting one packet into multiple pieces so that they can be arranged into different probabilistic. Stacking, on the other hand, combines multiple packets (or packet fragments) into one longer packet-size piece. For each flow mutation, we first sample the probabilistic difference between each $L_p$ and ${L_p}^`$. And based on the volume of differences, we decide if each $L_p$ bin needs to be fragmented or stacked. And then, we matched each bin probability based on the previously decided strategy. If there is any distribution we can mutate simply by fragmenting and stacking, then we use padding to fill the gap.

\begin{figure*}
        \centering
        \includegraphics[scale = 0.6]{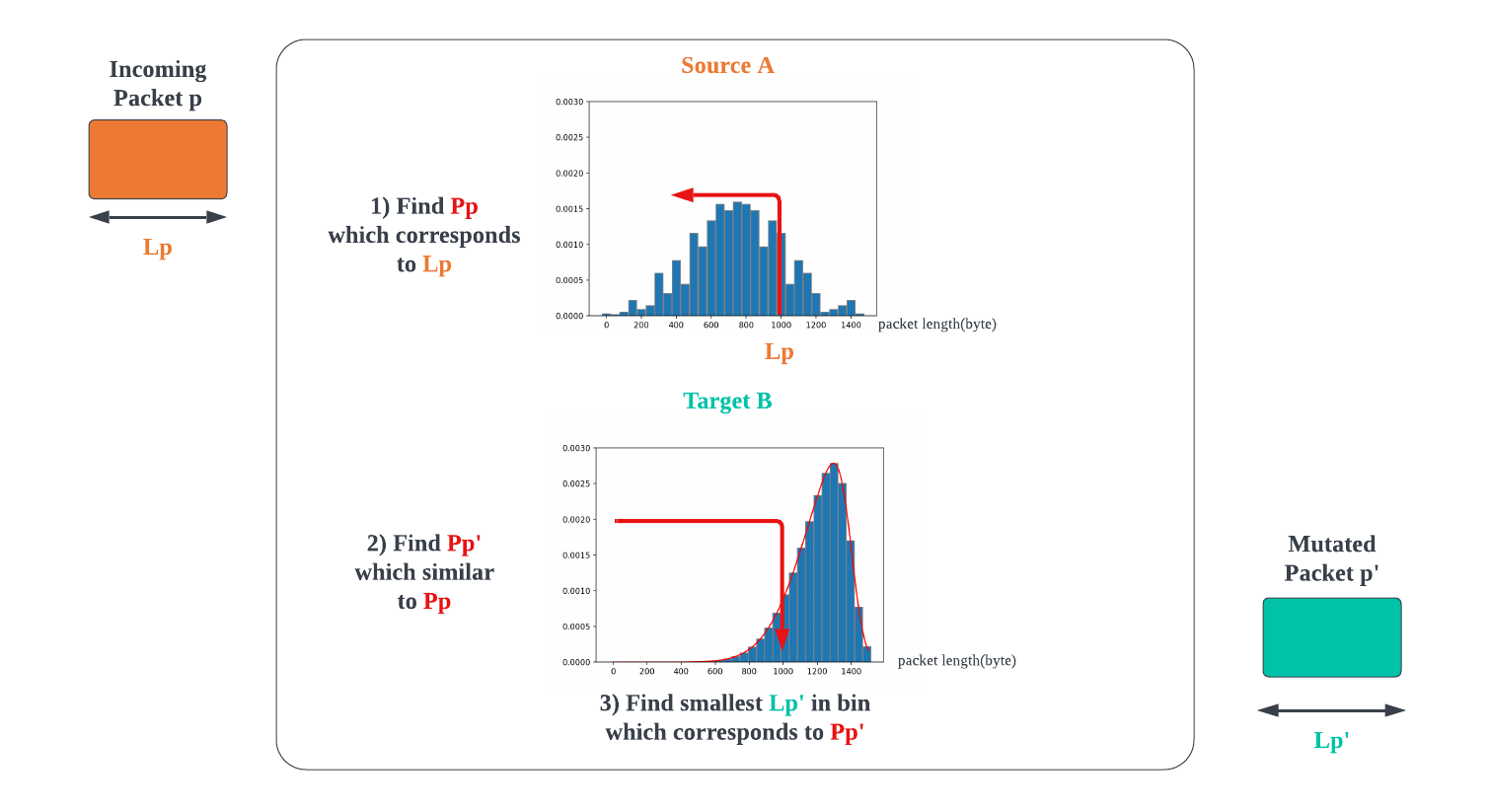}
        \caption{Packet Length Distribution Modification Pipeline}
        \label{fig: packet_modification}
\end{figure*}
\par
With the modification algorithm shown above (also fig.\ref{fig: packet_modification}), we essentially mutate the packet size probabilistic distribution from the source app to the mutated probabilities curve with the smallest overhead, which releases bandwidth for connection between end nodes. The desired mutation probabilities curve could be adjusted by setting new $\mu$, $\sigma$ ,and $\alpha$, and it would be beneficial to change the target distribution at each run to prevent the attacker from knowing the mapping of the source app to the specific curve.

\subsection{Experiment Setup and Evaluation Metric}
MIRAGE\cite{b25} is a reproducible architecture for mobile-app traffic capture and ground-truth creation. In our paper, we use an outcome of this system, MIRAGE-2019, a human-generated dataset for mobile traffic analysis combined with an AppScanner program to evaluate our algorithm.
\par
In the following section, we will discuss separately the statistical characteristics for MIRAGE-2019 and the details of AppScanner processing. The experiment evaluation metric will also be discussed.

\subsubsection{Statistic characteristics for MIRAGE-2019 dataset}
\par
The MIRAGE-2019 dataset gathers the traffic generated by 40 Android apps belonging to 16 different categories
according to Google Play apps distribution portal\cite{b26}. The released MIRAGE-2019 is a JSON formatted dataset in which one JSON file corresponds to one PCAP trace captured. And for each bitflow, the data collector extracts three feature groups-- per-packet data, per-flow features, and per-flow metadata.
\par
In the per-packet data (shown in TABLE \ref{table:1}), it contains six informative header fields including (1) source transport-layer, (2) port destination transport-layer port, (3) packet direction (0 for upstream, 1 for downstream), (4) IAT, (5) TCP window size (0 for UDP packets) and (6) number of bytes in L4 payload, also with the byte-wise raw L4 payload extracted from the first 32 packets of each bitflow.

\begin{table}
\centering  
\caption{Per-packet data structure}
\label{per-packet}
\begin{tabular}{|c|l|}
\hline
\textbf{Keys}               & \multicolumn{1}{c|}{\textbf{Description}} \\ \hline
\textit{src\_port}          & Source transport-layer port               \\ \hline
\textit{dst\_port}          & Destination transport-layer port          \\ \hline
\textit{packet\_dir}        & Packet direction                          \\ \hline
\textit{L4\_payload\_bytes} & Number of bytes in L4 payload             \\ \hline
\textit{iat}                & Inter-arrival time                        \\ \hline
\textit{TCP\_win\_size}     & TCP window size                           \\ \hline
\textit{L4\_raw\_payload}   & Byte-wise raw L4 payload                  \\ \hline
\end{tabular}
\label{table:1}
\end{table}

\par
The per-flow features (shown in Table \ref{table:2}) provide information on the whole bitflow and corresponding upstream and downstream flows, including 17 statistical features like minimum and maximum computed on the sets of upstream, downstream, and complete IP packet lengths and inter-arrival times.

\begin{table}
\centering  
\caption{Per-flow features structure}
\begin{tabular}{|c|l|}
\hline
\textbf{Keys}                   & \multicolumn{1}{c|}{\textbf{Description}} \\ \hline
\textit{min}                    & Minimum                                   \\ \hline
\textit{max}                    & Maximum                                   \\ \hline
\textit{mean}                   & Arithmetic mean                           \\ \hline
\textit{std}                    & Standard deviation                        \\ \hline
\textit{var}                    & Variance                                  \\ \hline
\textit{mad}                    & Mean absolute deviation                   \\ \hline
\textit{skew}                   & Unbiased sample skewness                  \\ \hline
\textit{kurtosis}               & Unbiased Fisher kurtosis                  \\ \hline
\textit{10$\sim$90\_percentile} & Percentile per 10                         \\ \hline
\end{tabular}
\label{table:2}
\end{table}

\par
The per-flow metadata (shown in Table \ref{table:3}) is information sets related to the complete bitflow and upstream and downstream flows, including Android-packet name, exact or most-common labeling, the number of packets, total bytes in IP packets, total bytes in L4 payloads and also flow duration in seconds.

\begin{table}
\centering
\caption{Per-flow metadata structure}
\begin{tabular}{|c|l|}
\hline
\textbf{Sub keys}                     & \multicolumn{1}{c|}{\textbf{Description}} \\ \hline
\textit{BF\_label}                    & Android-package name                      \\ \hline
\textit{BF\_labeling\_type}           & Exact or most-common labeling             \\ \hline
\textit{BF/UF/DF\_num\_packets}       & Number of packets                         \\ \hline
\textit{BF/UF/DF\_IP\_packet\_bytes}  & Total bytes in IP packets                 \\ \hline
\textit{BF/UF/DF\_L4\_payload\_bytes} & Total bytes in L4 payloads                \\ \hline
\textit{BF/UF/DF\_duration}           & Flow duration in seconds                  \\ \hline
\end{tabular}
\label{table:3}
\end{table}

A total of 4606 PCAP traces are included in the MIRAGE-2019 dataset, which is updated over time. Because the amount of data is large enough and the information it offers covers all the elements we require, we use it as our experimental data. Additionally, the mobile traffic data is generated by three separate devices, and more than 280 experimenters participated in the experimental sessions, ensuring that the data bias is reduced\cite{b25}.

\subsubsection{AppScanner}
\par
The AppScanner program is a classifier for recognizing applications with network traffic. The code is adapted from the partial implementation of FlowPrint\cite{b27}, which implements the Single Large Random Forest Classifier of AppScanner\cite{b28}. When testing traffic data from MIRAGE2019 with AppScanner, we extract six features from the per-packet features group for each bitflow, where each packet is represented as a list of (1) timestamp, (2) source IP, (3) destination IP, (4) TCP source port, (5) TCP destination port and (6) packet length.
\par
In our attack model, the attackers are only aware of the packet flow patterns, such as packet size, source/destination IP, and source/destination port. In order to classify the applications that users installed, the attackers examine the side channel data (IP packet headers) that is sent together with the encrypted data flow. Our proposed mobile-app traffic mutation algorithm may be more effectively evaluated because the AppScanner we are applying in the experiments similarly achieves the classification results using the same packet flow information as the attacker model.

\subsubsection{Evaluation Metric}
\par
In this experiment, we measure two metrics to evaluate the performance of our obfuscation algorithm. The first one is classification accuracy. For the traffic capture for each application in MIRAGE-2019, we calculate the number of times for correct labeling $n_{correct}$ versus the total labeling times $n_{total}$ from AppScanner, which is derived as:
\begin{equation}
\label{eqn acc}
    \text{\% Accuracy }= {{n_{correct}}\over {n_{total}}}
\end{equation}

\par
The second metric we measure for each test experiment is the overhead for each packet flow. We could be represented as following equation
\begin{equation}
\label{eqn9}
            \text{\% overhead }= {{L_{mutate}-L_{origin}}\over {L_{origin}}}
\end{equation}
where $L_{mutate}$ represents the total packet size after the traffic being mutated; and $L_{origin}$ represents the original total packet size before editing.

\subsection{Algorithm parameters}
In the experiment, we set our $alpha$ as a time variable of timestamp $t$, where 
\begin{equation}
\label{eqn6}
     \alpha = t\mod 60.0 \times (-1)^t
\end{equation}
And the mean of the desired distribution is calculated with
\begin{equation}
\label{eqn7}
     \mu = {1\over {n_{packet}}} {\sum_{x=0}^{{n_{packet}}} Length_{packet_n}}
\end{equation}
where ${n_{packet}}$ is number of packets that we going to mutate.
To cover 99.9\% of desired distribution, we set our standard deviation with
\begin{equation}
\label{eqn8}
     \sigma = {\mu \over 3}
\end{equation}

\section{Empirical Result}

\par
To evaluate the performance of our obfuscation algorithm, we first train a machine learning model from AppScanner for each application traffic data from MIRAGE-2019. We implement all six features for each packet flow provided from the dataset. And we divide the data into 75\% for the training and 25\% for testing, where the division is equally applied for each application. For validation, we use 5-fold cross-validation to evaluate the training model. The classification accuracy for each application is shown in Table.\ref{table:3}. We can see that most of the classifications achieve relatively high accuracy, except the accuracy drops when the AppScanner classifies some of the applications, which might be caused by over-fitting training.

\par
And then, we mutated the origin 25\% testing traffic data with our obfuscation algorithm mentioned in Section \ref{framworks}. For each application, we measure the classification accuracy from AppScanner, and we also calculate the overhead of packet flow for each application, and the result is shown in Table.\ref{table:4}. For accuracy, while classification for most of the application traffic data is 
dropped down to roughly 20\%, we still see there are several traffic data still remain a relatively high accuracy of classification. One guess could be these apps' traffic generally remains the same packet size for most of the flows, which makes the algorithm hard to obfuscate. Also, the data formation can also be one of the affected factors since the fixed form of the data structure flow would also reveal some pattern of the application feature. For the overhead, there is also a trend that there is an increasing overhead size while the packet size of application traffic stays relatively the same all the time. But for the other application traffic, the overhead keeps at an average 15\% which is a bit higher than the peer work. But since our obfuscation model is generated with timestamp instead of choosing specific application data distribution, the performance is evaluated as showing a great result.

\begin{table}[h]
\caption{Classification Accuracy for Application original traffic}
\centering
\begin{adjustbox}{width=\linewidth}

\begin{tabular}{llc}
\hline
Application Traffic        & Category           & \multicolumn{1}{l}{\% AppScanner Classification Accuracy}  \\ \hline
Pinteret origin            & Social             & 92.4                              \\
Facebook origin            & Social             & 93.5                              \\
Spotify origin             & Music and Audio    & 96.4                              \\
Wish origin                & Shopping           & 87.6                              \\
Groupon origin             & Shopping           & 90.1                              \\
TripAdvisor origin         & Travel and Local   & 94.5                              \\
Dropbox  origin            & Productivity       & 88.6                              \\
Trello origin              & Productivity       & 87.5                              \\
Viber origin               & Communication      & 93.7                              \\
Messenger origin           & Communication      & 91.5                              \\
Twitter origin             & News and Magazines & 92.7                              \\
Youtube origin             & Video Players      & 89.4                              \\
OneFootball origin         & Sports             & 91.5                              \\
AccuWeather origin         & Weather            & 94.1                              \\
Comics origin              & Comics             & 94.5                              \\
FourSquare origin          & Travel and Local   & 90.2                              \\
Subito origin              & Lifestyle          & 92.3                              \\
Duolingo origin            & Education          & 89.4                              \\
Waze origin                & Maps \& Navigation & 94.2                              \\
Slither.io origin          & Games              & 96.4                              \\ \hline
\end{tabular}
\end{adjustbox}
\label{table: 4}
\end{table}

\begin{table}
\caption{Classification Accuracy for Application mutated traffic}
\centering
\begin{adjustbox}{width=\linewidth}
\begin{tabular}{llcc}
\hline
Application Traffic         & Category           & \multicolumn{1}{l}{\% AppScanner Classification Accuracy} & \multicolumn{1}{l}{\% Overhead} \\ \hline
Pinteret mutated            & Social             & 21.6                             & 13.4         \\
Facebook mutated            & Social             & 16.4                             & 11.2         \\
Spotify mutated             & Music and Audio    & 46.5                             & 35.7         \\
Wish mutated                & Shopping           & 15.9                             & 10.6         \\
Groupon mutated             & Shopping           & 11.7                             & 12.8         \\
TripAdvisor mutated         & Travel and Local   & 15.5                             & 14.2         \\
Dropbox mutated             & Productivity       & 45.3                             & 26.4         \\
Trello mutated              & Productivity       & 56.1                             & 27.6         \\
Viber mutated               & Communication      & 9.7                              & 11.2         \\
Messenger mutated           & Communication      & 8.4                              & 9.7          \\
Twitter mutated             & News and Magazines & 24.6                             & 14.5         \\
Youtube mutated             & Video Players      & 35.4                             & 28.1         \\
OneFootball mutated         & Sports             & 13.4                             & 13.7         \\
AccuWeather mutated         & Weather            & 17.7                             & 15.4         \\
Comics mutated              & Comics             & 26.3                             & 24.5         \\
FourSquare mutated          & Travel and Local   & 17.3                             & 12.7         \\
Subito mutated              & Lifestyle          & 6.7                              & 10.4         \\
Duolingo mutated            & Education          & 16.8                             & 11.3        \\
Waze mutated                & Maps \& Navigation & 29.2                             & 17.4        \\
Slither.io mutated          & Games              & 23.4                             & 19.3        \\ \hline
\end{tabular}
\end{adjustbox}
\label{table:4}
\end{table}

\section{Conclusion}
With the increasing explosion of network-based applications nowadays, a massive volume of data traffic is being generated, and the data collector may infer a lot more information from that data traffic by knowing the types of applications a user has installed. Numerous different encryption techniques have been used to maintain the confidentiality of traffic in order to preserve users' privacy. However, the old methods of data encryption have been weakened by the increase in traffic classification done with machine learning methodology. In order to better secure users' privacy, we propose a new mobile-app traffic mutation algorithm in our research project.
\par
In this paper, we proposed a practical method for mobile application traffic obfuscation. We formalize the mathematical model for obfuscation probabilistic distribution, which allows the application mutates its traffic data without pre-mapping other types of application traffic flow size distribution to achieve a relatively good performance. The accuracy drops at least 50\% in the selected application traffic from MIRAGE2019. And the overhead keeps at a level which the highest is less than 40\%. The algorithm could have strong help in preventing behavior analysis based on only the traffic flow feature analysis and classification.
\par
In the future, we plan to improve the performance of our algorithm on applications that shows a homogeneous trend of packet size. And we also plan to implement noise factor into the model, so the computation cost can be decreased further.

\vspace{12pt}

\end{document}